*Article*

# High-production-rate fabrication of low-loss lithium niobate electro-optic modulators using photolithography assisted chemo-mechanical etching (PLACE)

**Rongbo Wu** [1,2], **Lang Gao** [1,2], **Youting Liang** [3,4], **Yong Zheng** [3,4], **Junxia Zhou** [3,4], **Hongxin Qi** [3,4,7], **Difeng Yin** [1,2], **Min Wang** [3,4], **Zhiwei Fang** [3,4], and **Ya Cheng** [1,2,3,4,5,6,8]

[1] State Key Laboratory of High Field Laser Physics, Shanghai Institute of Optics and Fine Mechanics, Chinese Academy of Sciences, Shanghai 201800, China
[2] University of Chinese Academy of Sciences, Beijing 100049, China
[3] State Key Laboratory of Precision Spectroscopy, East China Normal University, Shanghai 200062, China
[4] XXL—The Extreme Optoelectromechanics Laboratory, School of Physics and Electronics Science, East China Normal University, Shanghai 200241, China
[5] Collaborative Innovation Center of Light Manipulations and Applications, Shandong Normal University, Jinan 250358, China
[6] Shanghai Research Center for Quantum Sciences, Shanghai 201315, China
[7] Correspondence: hxqi@phy.ecnu.edu.cn
[8] Correspondence: ya.cheng@siom.ac.cn

**Abstract:** Integrated thin-film lithium niobate (LN) electro-optic (EO) modulators of broad bandwidth, low insertion loss, low cost and high production rate are essential elements in contemporary interconnection industries and disruptive applications. Here, we demonstrated the design and fabrication of a high performance thin-film LN EO modulator using photolithography assisted chemo-mechanical etching (PLACE) technology. Our device shows a 3-dB bandwidth over 50 GHz, along with a comparable low half wave voltage-length product of 2.16 V·cm. We obtain a fiber-to-fiber insertion loss of 2.6 dB.

## 1. Introduction

Integrated Mach-Zehnder modulators of high light-modulation rates, low power consumption and small sizes are essential elements in contemporary interconnection industries. They are also promising candidates to serve as the building blocks for disruptive applications ranging from quantum information processing and microwave photonics to artificial neural networks [1–6]. To meet the stringent requirements of industrial standard including a low optical insertion loss, low drive voltage, broad bandwidth, dense integration, robustness, low cost and high production rates, various material platforms and fabrication approaches have been investigated. For instance, silicon, lithium niobate (LN), indium phosphide, polymers and plasmonic have been explored to construct highly integrated light modulators of low production cost [7–10]. Notably, thin film LN (TFLN), which is a relatively new optical material featuring a high electro-optic coefficient, a large optical nonlinearity, and a broad transmission window, has attracted significant attention over the past decade [11–22]. Unlike the conventional bulk LN electro-optical modulators widely used in nowadays optical communication industry, integrated photonic circuits (PICs) built upon the TFLN can readily break up the bottlenecks of the bulk LN modulators in terms of the device size, drive voltage and bandwidth. On the other hand, with the shrinking sizes of the optical mode profiles in the ridge waveguides formed on the TFLN substrate, challenges have been encountered in the high-production-rate fabrication of low-loss, large-scale TFLN PIC devices. Electron beam writing can provide sufficiently high fabrication precision whilst suffers from a relatively low production yield. Ultraviolet (UV) lithography technology can provide high fabrication efficiency, whilst uncertainty still exists in terms of uniformity of wafer-scale production and propagation loss induced by the sidewall roughness.

To address the challenges, photolithography assisted chemo-mechanical etching (PLACE) technology has emerged as a promising technology for thin-film LN device production. During the last few years, various micro photonic components fabricated using PLACE technology including a microresonator with quality up to $10^8$ [23], waveguides with propagation loss as low as 0.027 dB/cm [24], a waveguide delay lines with meter scale length [25] and a waveguide amplifier with net gain over 20 dB [26] have been reported. The PLACE technology can support large device footprint, high fabrication uniformity, and competitive production rate simultaneously thanks to the high average power femtosecond laser and high-speed large-motion-range position stages. PLACE also gives rise to smooth waveguide sidewall and thus provides low device optical loss.

Here we demonstrate high performance thin-film LN EO modulators fabricated using PLACE technology. Our device shows a 3-dB electro-optic bandwidth over 50 GHz, along with a half wave voltage-length product ($V_\pi L$) of 2.16 V·cm. We obtain a fiber-to-fiber insertion loss of 2.6 dB.

## 2. Simulations and Device Design

Figure 1(a) shows the simulated optical field of the fundamental TE mode for a typical LN ridge waveguide fabricated by PLACE technology. The electrical field distribution in the waveguide with a 1 V static voltage applied between the signal and ground electrodes is shown in figure 2(b). To optimize the optical propagation loss caused by the absorption of the electrodes with a reasonable half wave voltage-length production ($V_\pi L$), we calculated both the $V_\pi L$ and the optical propagation loss for different electrode-pair gaps and etching depth, as shown in figure 1(c). Considering the radio-frequency (RF) loss, the alignment accuracy of the lift off process, and the fabrication salability of the PLACE process, we choose an electrode-pair gap of 5.5 μm and a waveguide etching depth of 0.21 μm, which correspond to a theoretical $V_\pi L$ of 2.2 V·cm and an optical loss below 0.1 dB/cm.

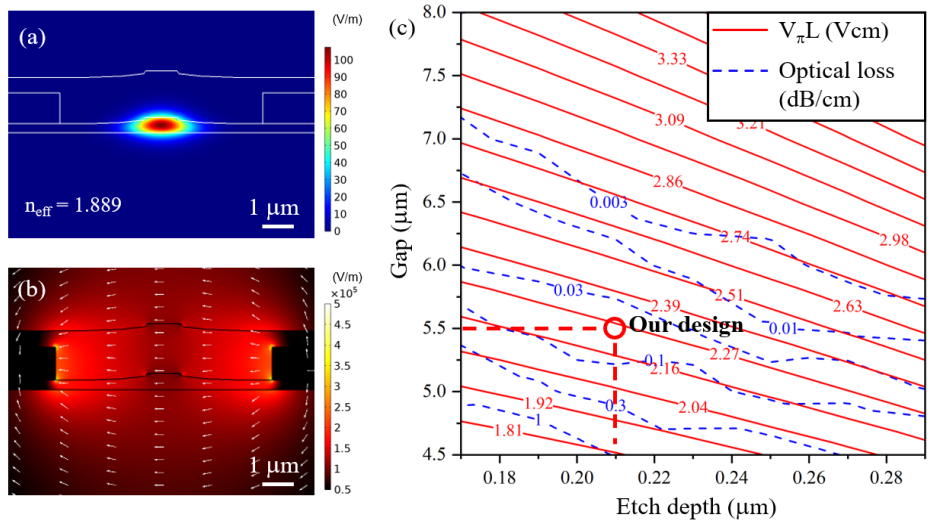

**Figure 1.** (a) Simulated optical field profile of the fundamental TE mode in the LN ridge waveguide. (b) Simulated static electric field when applying a 1 V voltage between the electrodes. (c) Contour map of both $V_\pi$ and optical loss versus different gaps and etching depths.

The design of the EO LN modulator is illustrated in Figure 2(a). The modulators are manufactured upon a commercial x-cut LNOI wafer (NANOLN) with a thin-film LN thickness of 500 nm. The modulator consists of two waveguide arms connected by a beam splitter and a combiner based on multimode interference (MMI) couplers. The electrodes carrying the RF signals propagating along the optical waveguide arms are designed and fabricated with the ground-signal-ground (GSG) configuration. Two groups of terminal pads with characteristic impedance of 50 Ω are fabricated on both the endss of the coplanar waveguide (CPW) transmission lines. The cross section of the modulator is shown in Figure 2(b). The GSG electrodes are arranged on the same plane of the waveguide arms to achieve better EO interaction. A silicon oxynitride (SiON) layer is deposited on top of the fabricated waveguides and the electrodes to obtain a better matching between the group velocity of optical field and RF field. Meanwhile, the SiON layer also functions as the cladding waveguide for the LN inverse tapers fabricated at both the ends of the modulator which together form the spot size converter (SSCs). The SSCs provides high optical coupling efficiency between the fabricated modulator and ultra-high numerical aperture fibers, giving rise to a favorable low fiber-to-fiber optical insertion loss.

Based on the geometric configuration in Figure 1, the group velocity matching between the RF field on the travelling electrode and the TE mode optic field in the ridge waveguide is numerically analyzed. We plot the simulation results in Figure 2(c). As shown by the curve of group refractive index versus frequency, the group velocity of the RF signals matches nicely with that of the optical field (up to a band of 50 GHz), which has been verified by our experimental measurement as discussed below (see Figure 5(b)). Also, we present the simulated characteristic impedance of the travelling electrode in Figure 2(c). It can be seen that the characteristic impedance does not change much (<4 Ω) with the varying frequency, which indicates a nearly perfect matching between the electrodes and terminal pads. This has also been confirmed by the experimental result as presented later in Figure 5(a). (See the curve of S11 vs. frequency in the figure).

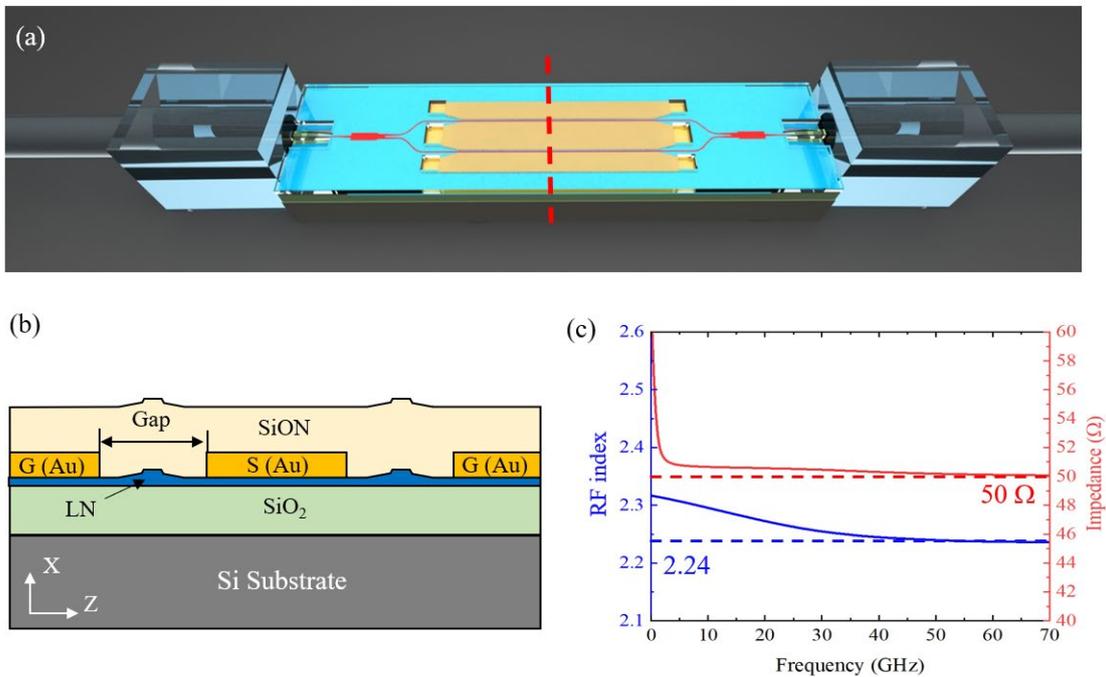

**Figure 2.** (a) Picture of the schematic TFLN modulator, UHNA fiber arrays are placed on both sides to couple light into and out of the chip. (b) Cross section of the modulation area. (c) The simulated RF group index and characteristic impedance of the modulator.

## 3. Measurement of Mode Profile in the Waveguide

Figure 3 (a) shows a photograph of a fabricated modulator chip. Fiber arrays were bonded on both ends of the chip using refractive index matching ultraviolet glue. The microscope image of the EO modulator is shown in figure 3(b). Figure 3(c) shows the SEM image of the cross section of the fabricated thin-film LN waveguide. Figure 3(d) shows the SEM image of the SSC, which can provide adiabatic conversion of the sub-micrometer mode area in the LN waveguide to micrometer mode area in the SiON waveguide. The mode field images captured by an infrared CCD are shown in figure 3(e) and (f) for both the SiON waveguide and UHNA fiber (UHNA7, CORNING), respectively. As can be seen from the images, the SSC offers favorable mode field overlap between the LN waveguide and the UHNA fiber.

To characterize the coupling loss between the SSC and the UHNA fiber, we coupled UHNA fibers with a straight LN waveguide of 1 mm and obtained a fiber-to-fiber insertion loss of ~2.0 dB. Since the propagation loss of such short waveguide can be negligible, the coupling loss between the SSC and the UHNA fibers can be estimated to be ~1.0 dB/facet. The fiber-to-fiber insertion loss of the fabricated EO modulator device was measured to be ~2.6 dB, including a coupling loss of 1.0 dB/facet and an on-chip propagation loss of ~0.6 dB.

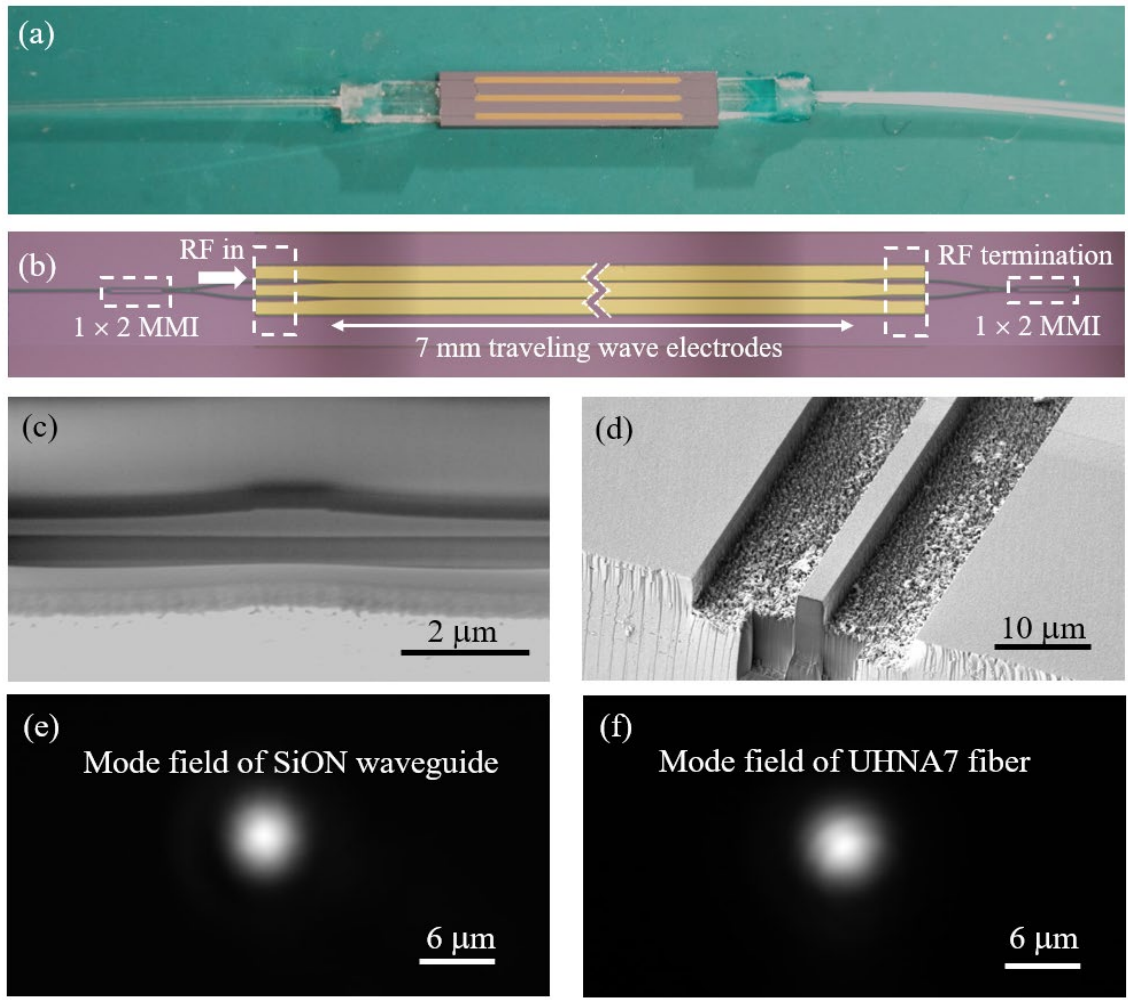

Figure 3 (a) Photograph of a fabricated EO modulator bonded with fiber arrays. (b) The microscope image of the EO modulator part of the chip. (c) The SEM image of the cross section of the fabricated thin-film LN waveguide. (d) The SEM image of the SiON waveguide. (e) The mode field profile of the SiON waveguide. (f) The mode field profile of the commercial UHNA7 fiber.

**4. Performance characterization of the fabricated device**

The schematic measurement set-up for the broadband EO response characterization of the device under test (DUT) is shown in Figure 4. A laser beam at 1550 nm wavelength was first amplified using an erbium-doped fiber amplifier (EDFA) and then sent into an inline fiber polarization controller (FPC) to generate the TE polarization mode as the light source of the DUT. The light was coupled into and out of the DUT using UHNA fibers. An RF signal with a frequency sweep continuously from 10 MHz up to 50 GHz was provided by a vector network analyzer (VNA, R&S ZNA 50) which was applied on the DUT through a 50 GHz GSG probe (GGB, model 50A). An additional probe was used to connect the DUT with an external 50 Ω terminator to reduce the RF signal reflection caused by impedance mismatching. The modulated light was converted to RF signal by a photodiode (PD, New Focus 1014) and recorded by the VNA to analyze the S parameters.

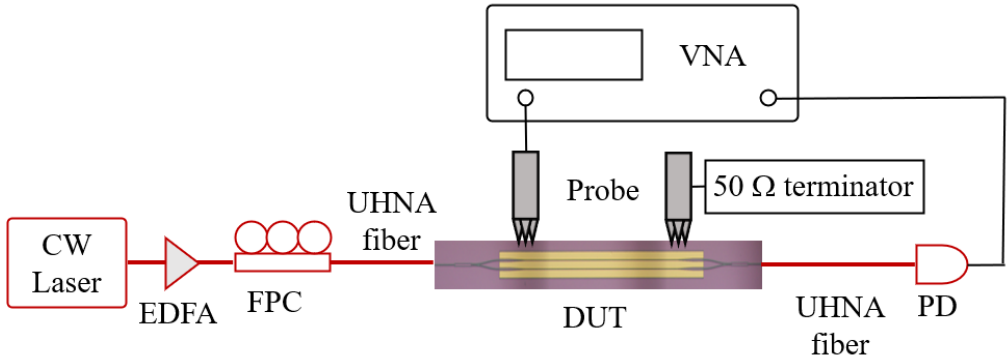

**Figure 4.** The schematic view of the measurement set-up for broadband EO response characterization. EDFA, erbium doped fiber amplifier; DUT, device under test; FPC, fiber polarization controller; PD, photodiode; VNA, vector network analyzer.

To verify the calculation result of the RF field, we further measured the broadband electrical-electrical (EE) response of the electrodes on the fabricated modulator. Figure 5(a) shows the EE transmission (S21) and reflection (S11) parameter for the modulator with 7 mm long arm length. The roll off of the S21 curve is 4.9 dB, implies an acceptable RF loss, since the 3 dB EO bandwidth can be estimated by the 6.41 dB EE bandwidth in the case of perfectly velocity matching [18]. The S21 curve maintains below -30 dB, implies a good matching of the characteristic impedance $Z_0$ between the CPW electrodes and the probes. The extracted $Z_0$ together with the extracted RF group index verses the applied RF frequency are shown in Figure 5(b), indicating that the RF group index is close to the optical group index with a difference below 0.05.

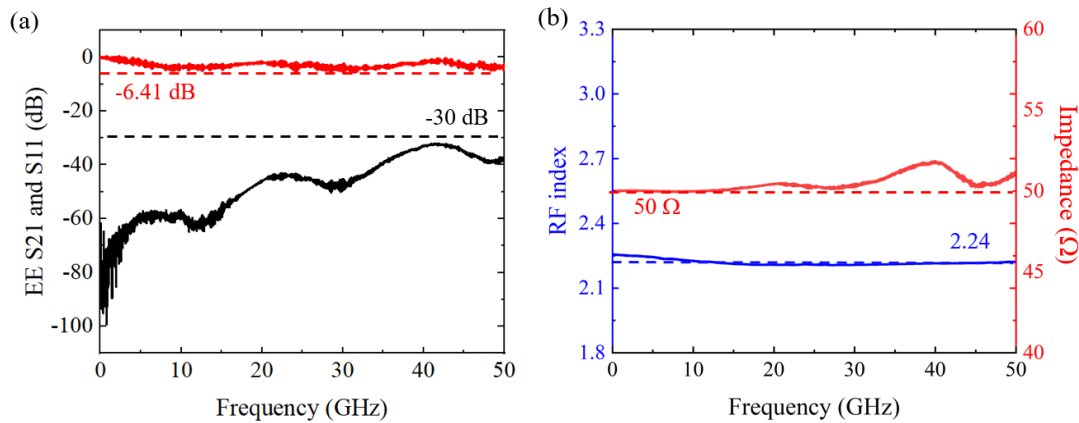

Figure 5 (a) The EE transmission (S21) and reflection (S11) parameter for the 7-mm-long modulator. (b) The extracted RF group index curve and characteristic impedance curve of the CPW electrodes.

To characterize the static electro-optical (EO) property of the fabricated modulators, a 100 kHz triangle wave signal with a peak-to-peak velocity of 20 V was applied on the modulator. The EO responses of the 7 mm long modulator is shown in Figure 5(a) and (b), from which the $V_\pi$ was measured to be 3.1 V. The extinction ratio was measured to be -18 dB, thus the extracted $V_\pi L$ can be calculated as 2.16 V·cm. We then measured the broadband EO response of the fabricated modulators. The EO S21 measured as a function of the applied RF frequency from 10 MHz to 50 GHz is shown in Figure 6(c). One can see that the actual 3-dB bandwidth of the modulator is beyond 50 GHz, which is limited by the performance of our measurement equipment.

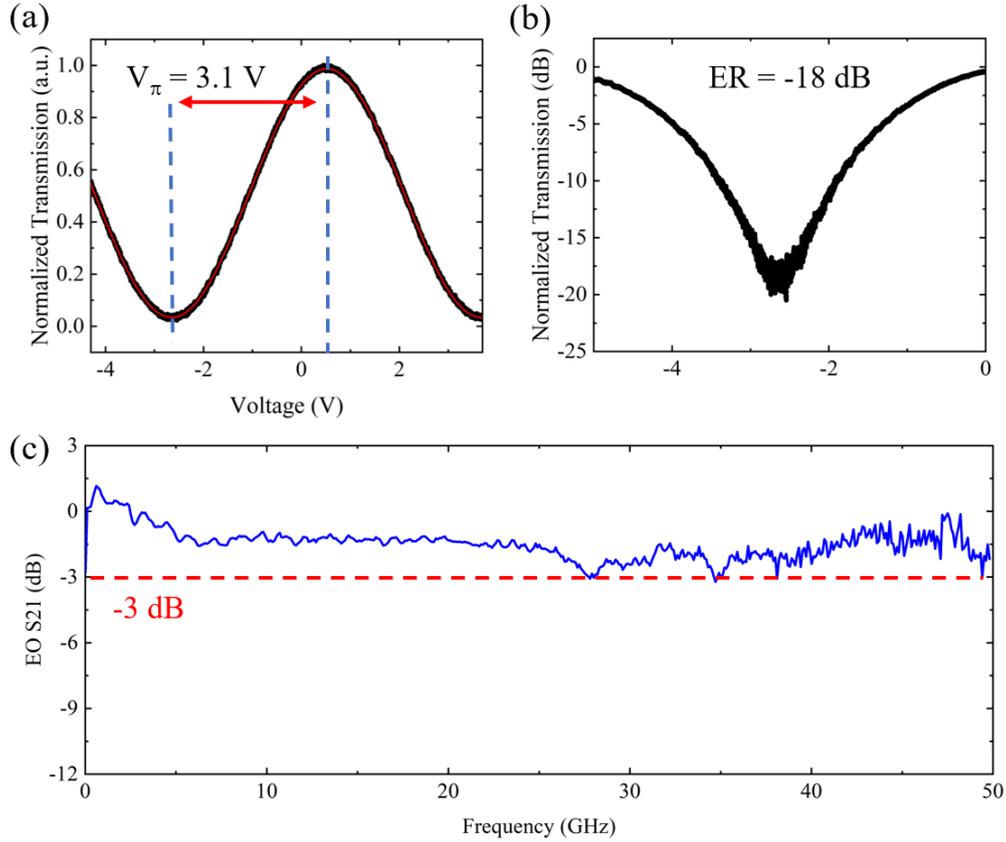

Figure 6 (a) The static EO response of the modulator. (b) The EO S21 curve up to 50 GHz of the modulator.

## 5. Conclusions

In conclusion, we have shown that high performance EO modulators featuring a low fiber-to-fiber insertion loss below 3 dB, a low $V_\pi$ of 3.1 V, and a high EO bandwidth over 50 GHz have been fabricated using the PLACE technique. Specifically, it takes only ~3 min to complete the mask patterning on the Cr film for each modulator using the femtosecond laser direct writing, which allows to produce the EO modulators of high yield and low cost. Benefitted from the low propagation loss of the LN waveguide and large-area fabrication capability of PLACE technology, our result shows promising potential for developing large-scale low-loss cascaded EO modulators.

**Funding:** This research was funded by the National Key R&D Program of China (2019YFA0705000), National Natural Science Foundation of China (Grant Nos. 12004116, 12104159, 11874154, 11734009, 11933005, 12134001, 61991444), Science and Technology Commission of Shanghai Municipality (NO.21DZ1101500), Shanghai Municipal Science and Technology Major Project (Grant No.2019SHZDZX01), Shanghai Sailing Program (21YF1410400).

**Conflicts of Interest:** The authors declare no conflict of interest.

## References

1. Cheng, Q.; Bahadori, M.; Glick, M.; Rumley, S.; Bergman, K. Recent Advances in Optical Technologies for Data Centers: A Review. *Optica* **2018**, *5*, 1354–1370.
2. Larger, L.; Soriano, M.C.; Brunner, D.; Appeltant, L.; Gutiérrez, J.M.; Pesquera, L.; Mirasso, C.R.; Fischer, I. Photonic Information Processing beyond Turing: An Optoelectronic Implementation of Reservoir Computing. *Optics express* **2012**, *20*, 3241–3249.
3. Marpaung, D.; Yao, J.; Capmany, J. Integrated Microwave Photonics. *Nature Photon* **2019**, *13*, 80–90, doi:10.1038/s41566-018-0310-5.
4. Fortier, T.M.; Kirchner, M.S.; Quinlan, F.; Taylor, J.; Bergquist, J.C.; Rosenband, T.; Lemke, N.; Ludlow, A.; Jiang, Y.; Oates, C.W. Generation of Ultrastable Microwaves via Optical Frequency Division. *Nature Photonics* **2011**, *5*, 425–429.
5. Sharma, S.; Eiswirt, P.; Petter, J. Electro Optic Sensor for High Precision Absolute Distance Measurement Using Multiwavelength Interferometry. *Optics express* **2018**, *26*, 3443–3451.
6. Shen, Y.; Harris, N.C.; Skirlo, S.; Prabhu, M.; Baehr-Jones, T.; Hochberg, M.; Sun, X.; Zhao, S.; Larochelle, H.; Englund, D.; et al. Deep Learning with Coherent Nanophotonic Circuits. *Nature Photonics* **2017**, *11*, 441–446, doi:10.1038/nphoton.2017.93.
7. Ogiso, Y.; Ozaki, J.; Ueda, Y.; Kashio, N.; Kikuchi, N.; Yamada, E.; Tanobe, H.; Kanazawa, S.; Yamazaki, H.; Ohiso, Y. Over 67 GHz Bandwidth and 1.5 V Vπ InP-Based Optical IQ Modulator with Nipn Heterostructure. *Journal of lightwave technology* **2016**, *35*, 1450–1455.
8. Sun, C.; Wade, M.T.; Lee, Y.; Orcutt, J.S.; Alloatti, L.; Georgas, M.S.; Waterman, A.S.; Shainline, J.M.; Avizienis, R.R.; Lin, S. Single-Chip Microprocessor That Communicates Directly Using Light. *Nature* **2015**, *528*, 534–538.
9. Lee, M.; Katz, H.E.; Erben, C.; Gill, D.M.; Gopalan, P.; Heber, J.D.; McGee, D.J. Broadband Modulation of Light by Using an Electro-Optic Polymer. *Science* **2002**, *298*, 1401–1403.
10. Haffner, C.; Chelladurai, D.; Fedoryshyn, Y.; Josten, A.; Baeuerle, B.; Heni, W.; Watanabe, T.; Cui, T.; Cheng, B.; Saha, S. Low-Loss Plasmon-Assisted Electro-Optic Modulator. *Nature* **2018**, *556*, 483–486.
11. Kong, Y.; Bo, F.; Wang, W.; Zheng, D.; Liu, H.; Zhang, G.; Rupp, R.; Xu, J. Recent Progress in Lithium Niobate: Optical Damage, Defect Simulation, and on-Chip Devices. *Advanced Materials* **2020**, *32*, 1806452.
12. Jia, Y.; Wang, L.; Chen, F. Ion-Cut Lithium Niobate on Insulator Technology: Recent Advances and Perspectives. *Applied Physics Reviews* **2021**, *8*, 011307.
13. Qi, Y.; Li, Y. Integrated Lithium Niobate Photonics. *Nanophotonics* **2020**, *9*, 1287–1320, doi:10.1515/nanoph-2020-0013.
14. Wang, C.; Zhang, M.; Chen, X.; Bertrand, M.; Shams-Ansari, A.; Chandrasekhar, S.; Winzer, P.; Lončar, M. Integrated Lithium Niobate Electro-Optic Modulators Operating at CMOS-Compatible Voltages. *Nature* **2018**, *562*, 101–104, doi:10.1038/s41586-018-0551-y.
15. He, M.; Xu, M.; Ren, Y.; Jian, J.; Ruan, Z.; Xu, Y.; Gao, S.; Sun, S.; Wen, X.; Zhou, L.; et al. High-Performance Hybrid Silicon and Lithium Niobate Mach–Zehnder Modulators for 100 Gbit S−1 and Beyond. *Nature Photonics* **2019**, *13*, 359–364, doi:10.1038/s41566-019-0378-6.
16. Luke, K.; Kharel, P.; Reimer, C.; He, L.; Loncar, M.; Zhang, M. Wafer-Scale Low-Loss Lithium Niobate Photonic Integrated Circuits. *Optics Express* **2020**, *28*, 24452–24458.
17. Kharel, P.; Reimer, C.; Luke, K.; He, L.; Zhang, M. Breaking Voltage–Bandwidth Limits in Integrated Lithium Niobate Modulators Using Micro-Structured Electrodes. *Optica* **2021**, *8*, 357, doi:10.1364/OPTICA.416155.
18. Liu, Y.; Li, H.; Liu, J.; Tan, S.; Lu, Q.; Guo, W. Low $V_\pi$ Thin-Film Lithium Niobate Modulator Fabricated with Photolithography. *Opt. Express* **2021**, *29*, 6320, doi:10.1364/OE.414250.
19. Ying, P.; Tan, H.; Zhang, J.; He, M.; Xu, M.; Liu, X.; Ge, R.; Zhu, Y.; Liu, C.; Cai, X. Low-Loss Edge-Coupling Thin-Film Lithium Niobate Modulator with an Efficient Phase Shifter. *Opt. Lett.* **2021**, *46*, 1478, doi:10.1364/OL.418996.

20. Lin, J.; Bo, F.; Cheng, Y.; Xu, J. Advances in On-Chip Photonic Devices Based on Lithium Niobate on Insulator. *Photonics Research* **2020**, *8*, 1910–1936.
21. Yang, F.; Fang, X.; Chen, X.; Zhu, L.; Zhang, F.; Chen, Z.; Li, Y. Monolithic Thin Film Lithium Niobate Electro-Optic Modulator with over 110 GHz Bandwidth. *Chin. Opt. Lett.* **2022**, *20*, 022502, doi:10.3788/COL202220.022502.
22. Xu, M.; Zhu, Y.; Pittalà, F.; Tang, J.; He, M.; Ng, W.C.; Wang, J.; Ruan, Z.; Tang, X.; Kuschnerov, M.; et al. Dual-Polarization Thin-Film Lithium Niobate in-Phase Quadrature Modulators for Terabit-per-Second Transmission. *Optica* **2022**, *9*, 61, doi:10.1364/OPTICA.449691.
23. Gao, R.; Yao, N.; Guan, J.; Deng, L.; Lin, J.; Wang, M.; Qiao, L.; Fang, W.; Cheng, Y. Lithium Niobate Microring with Ultra-High Q Factor above 10 8. *Chinese Optics Letters* **2022**, *20*, 011902.
24. Wu, R.; Wang, M.; Xu, J.; Qi, J.; Chu, W.; Fang, Z.; Zhang, J.; Zhou, J.; Qiao, L.; Chai, Z. Long Low-Loss-Litium Niobate on Insulator Waveguides with Sub-Nanometer Surface Roughness. *Nanomaterials* **2018**, *8*, 910.
25. Zhou, J.; Gao, R.; Lin, J.; Wang, M.; Chu, W.; Li, W.; Yin, D.; Deng, L.; Fang, Z.; Zhang, J. Electro-Optically Switchable Optical True Delay Lines of Meter-Scale Lengths Fabricated on Lithium Niobate on Insulator Using Photolithography Assisted Chemo-Mechanical Etching. *Chinese Physics Letters* **2020**, *37*, 084201.
26. Liang, Y.; Zhou, J.; Liu, Z.; Zhang, H.; Fang, Z.; Zhou, Y.; Yin, D.; Lin, J.; Yu, J.; Wu, R.; et al. A High-Gain Cladded Waveguide Amplifier on Erbium Doped Thin-Film Lithium Niobate Fabricated Using Photolithography Assisted Chemo-Mechanical Etching. *Nanophotonics* **2022**, *0*, doi:10.1515/nanoph-2021-0737.